# Does cluster encapsulation inhibit sintering? Stabilization of size-selected Pt clusters on Fe$_3$O$_4$(001) by SMSI

*Sebastian Kaiser,[‡,1,2] Johanna Plansky,[‡,2,3] Matthias Krinninger,[2,3] Andrey Shavorskiy,[4] Suyun Zhu,[4] Ueli Heiz,[1,2] Friedrich Esch,[1,2,*] and Barbara A. J. Lechner [2,3,5,*]*

[1] Chair of Physical Chemistry, Department of Chemistry, School of Natural Sciences, Technical University of Munich, Lichtenbergstr. 4, 85748 Garching, Germany

[2] Catalysis Research Center, Technical University of Munich, Lichtenbergstr. 4, 85748 Garching, Germany

[3] Functional Nanomaterials Group, Department of Chemistry, School of Natural Sciences, Technical University of Munich, Lichtenbergstr. 4, 85748 Garching, Germany

[4] MAX IV Laboratory, Lund University, Lund 221 00 Sweden

[5] Institute for Advanced Study, Technical University of Munich, Lichtenbergstr. 2a, 85748 Garching, Germany

ABSTRACT The metastability of supported metal nanoparticles limits their application in heterogeneous catalysis at elevated temperatures due to their tendency to sinter. One strategy to overcome these thermodynamic limits on reducible oxide supports is encapsulation via strong metal-support interaction (SMSI). While annealing-induced encapsulation is a well-explored phenomenon for extended nanoparticles, it is as yet unknown whether the same mechanisms hold for sub-nanometer clusters, where concomitant sintering and alloying might play a significant role. In this article, we explore the encapsulation and stability of size-selected $Pt_5$, $Pt_{10}$ and $Pt_{19}$ clusters deposited on $Fe_3O_4(001)$. In a multimodal approach using temperature-programmed desorption (TPD), x-ray photoelectron spectroscopy (XPS) and scanning tunneling microscopy (STM), we demonstrate that SMSI indeed leads to the formation of a defective, FeO-like conglomerate encapsulating the clusters. By stepwise annealing up to 1023 K, we observe the succession of encapsulation, cluster coalescence and Ostwald ripening, resulting in square-shaped crystalline Pt particles, independent of the initial cluster sizes. The respective sintering onset temperatures scale with the cluster footprint and thus size. Remarkably, while small encapsulated clusters can still diffuse as a whole, atom detachment and thus Ostwald ripening are successfully suppressed up to 823 K, *i.e.* 200 K above the Hüttig temperature that indicates the thermodynamic stability limit.

1. INTRODUCTION

In order to maximize material efficiency and to fully exploit non-scalable electronic effects in very small, sub-nanometer particles (called clusters) as heterogeneous catalysts, one has to counteract the intrinsic tendency of these metastable particles to sinter. Typically, oxide supports provide adequate cluster binding sites at defects to prevent cluster diffusion, while dewetting on the stochiometric surface suppresses atom diffusion.[1] Additionally, stabilization of particles by overgrowth of thin reducible oxide layers through the strong metal-support interaction (SMSI) effect has recently come into the focus of interest again.[2] The concept, first discussed by Tauster et al.,[3] holds for group VIII metals on reducible oxides and is well-established in ultra-high vacuum (UHV) where the encapsulation layer, formed by reductive annealing, has been described as a reduced film of self-limiting thickness and defined stoichiometry for large nanoparticles and extended metal islands.[4,5] The stabilization of nanoparticles by SMSI effects has numerous implications in catalysis, ranging from methanol steam reforming,[6,7] over the water-gas shift reaction,[8] to low temperature CO oxidation,[9] and the hydrogen oxidation reaction (HOR) under harsh oxidative conditions.[10,11] Recent studies have shown that the growth process can be strongly dependent on pressure conditions,[12] and on steady state redox dynamics.[13] On the atomic scale, many questions still remain, particularly pertaining to the growth dynamics of the film, its permeability to diffusion of small reactants, and the stability of the encapsulated particles. To date, the encapsulation phenomenon by SMSI has only been investigated for nanoparticles, but its extension to the sub-nm scale is still an unexplored territory. Specifically, the small cluster footprint strongly facilitates cluster diffusion in the temperature regimes where encapsulation takes place. A fundamental understanding of the atomic-scale processes requires hence to disentangle encapsulation and sintering, in order to discern whether the clusters remain size-selected or might even be completely integrated into the substrate. In our research, we study the cluster encapsulation and concomitant sintering behavior on the atomic scale with sound statistics and across a wide temperature range, with the aim to enable heterogeneous catalysis at stabilized sub-nm particles under harsh conditions.

Clusters have been shown to exhibit various exciting size effects that can be exploited for their catalytic activity, in the extreme case leading to drastic changes in turnover frequencies with the addition or removal of just a single atom to/from the cluster.[14,15] As a consequence, where sintering control is already a key issue in nanoparticle catalysts, it is even more crucial when it comes to stabilizing specific cluster sizes. Two distinct sintering mechanisms have been described, Ostwald and Smoluchowski ripening.[1,16–18] Ostwald ripening is particle coarsening by detachment of single atoms, which diffuse and reattach to other particles.[19] While the maximum onset temperature depends on the particle material and can be estimated by the so-called Hüttig temperature, where atoms start to detach from undercoordinated sites,[20,21] smaller particles (i.e. those with a high curvature) such as clusters exhibit a higher vapor pressure and thus start to sinter much earlier.[22] This curvature dependence generally leads to particles of larger diameter growing at the expense of smaller ones. Smoluchowski ripening, on the other hand, is sintering by migration of entire particles and coalescence with neighboring ones, yielding only multiples of the original cluster size.[23] While Ostwald ripening critically depends on the atom detachment energy, the onset temperature for Smoluchowski ripening is determined by the strength of the particle-support interaction influencing the diffusion energy barrier, which scales approximately with the cluster footprint, typically resulting in an atom-by-atom size dependence.[24]

To date, various approaches to particle stabilization have been suggested. For sub-nanometer clusters, it has been shown that size selection alone can kinetically mitigate Ostwald ripening.[25]

However, theoretical calculations suggest that sintering can indeed be accelerated in cluster sizes that exhibit a large number of isomers that fluxionally interconvert.[26] Other stabilization methods such as alloying with stabilizing metals,[27,28] or increasing the particle distance by use of high surface area supports[29] have been shown to successfully prevent particle sintering. Last but not least, particle re-dispersion strategies have been presented as alternatives for reactivating catalysts.[30] Here, however, we follow the stabilization approach via SMSI as a pathway that maintains cluster size selection and hence the potential to fully control catalyst activity.

In the present work, we investigate the encapsulation and sintering of size-selected $Pt_{5-19}$ clusters deposited on a magnetite, $Fe_3O_4(001)$, support. Iron oxides are abundant materials that exhibit a rich redox chemistry,[31] and magnetite in particular is magnetic, reducible and conductive. It crystallizes in an inverse spinel structure, whereby the $O^{2-}$ anions form an fcc-lattice with $Fe^{2+}$ occupying octahedral sites, and $Fe^{3+}$ tetrahedral and octahedral sites in a 1:1 ratio.[32–34] The $Fe_3O_4(001)$ surface reconstructs into the subsurface cation vacancy (SCV) reconstruction, yielding a $(\sqrt{2} \times \sqrt{2})R45°$ structure with only $Fe^{3+}$ occupying the uppermost layers.[35] This reconstruction can be observed as parallel, undulating rows in scanning tunneling microscopy, which are rotated by 90° between two adjacent atomic terraces. The (001) surface has a large quantity of different defects,[36–39] which can participate in its surface chemistry as adsorption and dissociation sites for molecules.[40–43] Moreover, we have recently shown that lattice oxygen from the magnetite support can readily participate in catalytic reactions by migrating onto Pt clusters, which we termed lattice oxygen reverse spillover.[44] Here, we use temperature programmed desorption (TPD), high resolution synchrotron x-ray photoelectron spectroscopy (XPS) and scanning tunneling microscopy (STM) measurements to show the formation of an encapsulating SMSI layer even on these small clusters, investigate its redox state and morphology, and clarify the sintering mechanism for $Pt_5$, $Pt_{10}$ and $Pt_{19}$ clusters, where we pinpoint the cluster size-dependent onset of Smoluchowski ripening and subsequently that of Ostwald ripening.

2. EXPERIMENTAL METHODS

Natural $Fe_3O_4$(001) crystals (SurfaceNet GmbH), were prepared by a series of cleaning cycles, each consisting of $Ar^+$ ion bombardment (20 min, 4 x $10^{-5}$ mbar Ar, 1 keV, 5.0 µA sputter current) and subsequent annealing in an oxygen atmosphere (20 min, 5 x $10^{-7}$ mbar $O_2$, 983 K). The reproducibility of the preparation procedure as well as the cleanliness and stoichiometry of the crystals were checked regularly by means of STM and XPS. Size-selected Pt clusters were generated using a laser ablation cluster source.[45] Here, the 2$^{nd}$ harmonic of a Nd:YAG laser is used to evaporate Pt from a rotating target, yielding a plasma, which is subsequently cooled in the adiabatic expansion of a He pulse (Westfalen AG, grade 6.0), resulting in a broad distribution of clusters. A series of electrostatic lenses guides the clusters towards a 90° bender for charge selection, followed by a quadrupole mass filter for size selection. Finally, the clusters are deposited onto the substrate under soft landing conditions (kinetic energy < 1 eV/atom). For all experiments, cluster deposition was carried out at room temperature; a cluster coverage of 0.05 clusters/nm$^2$ was deposited.

The TPD and STM measurements were performed under ultrahigh vacuum conditions (UHV), with a system base pressure of < 1 x $10^{-10}$ mbar. Temperature control is performed with the same sample holder for all experiments, using a pyrolytic boron nitride heater located directly below the sample and a type-K thermocouple in direct contact with the sample, with an absolute estimated accuracy of ± 5 K. All TPD experiments were carried out using the sniffer, a mass spectrometer-based (Pfeiffer Vacuum GmbH, QMA 200 Prisma Plus) reactivity measurement device, designed for high sensitivity measurements of low coverage cluster samples.[46] To maximize the signal and ensure a low signal-to-noise ratio, all TPD-related samples have been saturated with $C^{18}O$ (Eurisotop, 96.1%) already during cluster deposition.

The STM measurements were performed in constant current mode with a commercial Scienta Omicron VT-AFM instrument, using home-made etched W-tips. In order to disentangle time and temperature effects, we systematically annealed the sample for 10 minutes in 50 K steps before measuring the resulting cluster distribution at room temperature. The STM images were processed with Gwyddion using the plane subtraction and row-by-row alignment tools for background correction.[47] The height distribution of the particles was determined using a home-written Igor routine, by detecting the particles via an intensity threshold, drawing a profile through the cluster maximum and determining the height of the cluster with respect to the median background of the image. For higher statistical significance, ten 100 x 100 nm² STM images were recorded both in the center and at the edge of the sample for each annealing temperature. A minority of clusters which were located at step edges, at the edge of the image or in small holes were excluded from the height analysis, but included in the coverage determination. In images with more than one terrace, a separate evaluation was carried out for each terrace.

High-resolution XPS measurements were performed at the APXPS endstation of the HIPPIE beamline,[48] MAX IV laboratory in Lund, Sweden, which is equipped with a Scienta Omicron HiPP-3 electron energy analyzer. A home-built UHV suitcase with a base pressure of approx. 1 x 10$^{-10}$ mbar was used for sample transfer after cluster deposition in our labs in Munich to the beamline in Lund, ensuring clean and intact samples. To desorb any adsorbates accumulated during the several day-long transport in the UHV suitcase (presumably mainly CO and traces of hydrocarbons), we started the experiment by heating the sample to 373 K to desorb the adsorbates.

This resulted in a shift of 0.6 eV to lower binding energies, corresponding to CO desorption,[49,50] as well as in a reduction of the peak width for the Pt 4f signal and no clear shift in the Fe 3p (compare Supporting Information Figure S1). For heating, an infrared laser directly illuminating the back of the sample was used; the temperature was measured with a type-K thermocouple mounted to the side of the crystal. Below, we indicate the temperatures as measured, which could be somewhat underestimated due to limited heat transfer. The XP spectra were acquired in UHV with photon energies of 307 eV and 921 eV and a constant pass energy and emission angle (normal emission). Initial checks showed no significant beam damage, thus we recorded all spectra at the same spot on the sample to allow quantitative comparison of the spectra. In data evaluation using KolXPD,[51] the binding energies were calibrated with respect to the Fermi edge and all spectra normalized to their low binding energy background. A Shirley background was subtracted from the spectra prior to quantitative analysis.

## 3. RESULTS AND DISCUSSION

*Encapsulation of sub-nm clusters induced by strong metal-support interaction.*

In order to investigate whether and how SMSI influences cluster sintering dynamics, we first need to establish that encapsulation of clusters actually occurs. In our recent work on lattice oxygen reverse spillover on Pt/Fe$_3$O$_4$(001), we have already found first indications for cluster encapsulation in our TPD experiments, namely the loss of CO adsorption sites and thus CO$_2$ formation capability upon heating beyond a certain temperature threshold.[44] We now investigate this phenomenon in more detail on the example of Pt$_{19}$, by means of TPD and high resolution XPS, focusing on the exact growth conditions, redox state and morphology of the encapsulating layer, starting with the CO TPD from our previous work (compare Figure 1, 1$^{st}$ run).

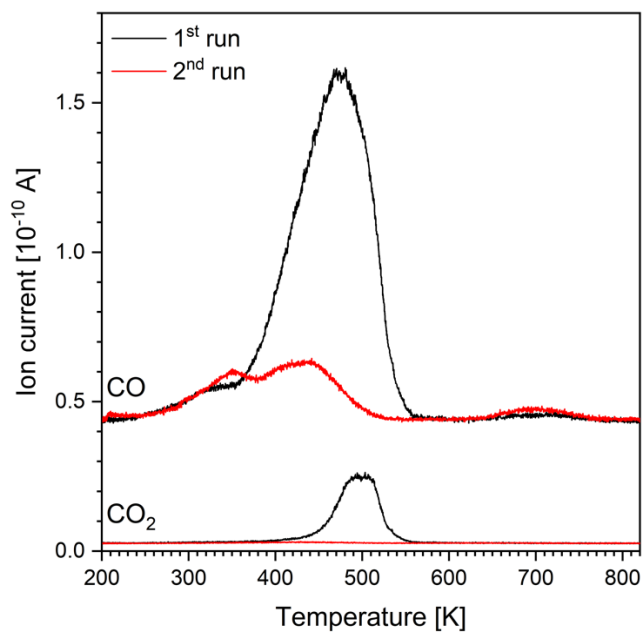

Figure 1. Two subsequent CO TPD spectra of $Pt_{19}$ clusters supported on $Fe_3O_4(001)$ recorded after exposing the sample to a saturating amount of $C^{18}O$ at 200 K (~10 L) before each run. The $C^{18}O$ (m/z = 30, top) and $C^{18}O^{16}O$ (m/z = 46, bottom) traces are shown. The heating rate was 1 K/s. A clear loss of the main desorption feature in the second run can be observed for both traces, with new, lower temperature features appearing in the CO trace, which hint at the presence of FeO defect sites.[52] Furthermore, the $CO_2$ production is completely suppressed in the 2nd run.

The main CO desorption feature in the 1st run (black), with a peak at 475 K, arises from desorption from the clusters, as it is not observed on the clean magnetite surface.[41,44] The rather weak, broad feature around 700 K, in contrast, belongs to recombinative desorption of CO from reduced Fe sites such as Fe adatoms.[37,53] In the 2nd TPD run (red), the main CO desorption peak has vanished completely and instead two new, overlapping and less intense desorption features with maxima at 350 K and 430 K emerge. Indeed, the new CO desorption features appearing in the second TPD are comparable to literature spectra from oxygen vacancy defect sites on FeO films.[52] Additionally, the desorption peak at 700 K has increased significantly in intensity compared to the first TPD run, again hinting at a more reduced surface which is in perfect agreement with an SMSI encapsulation state similar to that described for Pt nanoparticles on different magnetite facets.[5,54] Our STM studies showed that these TPD runs are ramped quickly enough that sintering is still avoided.[44]

Concurrently with the CO desorption, $CO_2$ formation is observed as a broad feature in the 1st TPD run, originating from catalytic CO oxidation with lattice oxygen,[44] which is completely lost in the 2nd run. When dosing the same quantity of clusters on top of the already heated sample, the same signal reappears in the 1st run after re-deposition and disappears again in a 2nd run (see Supporting Information Figure S2). Importantly, the resulting FeO defect-related peak intensity scales with Pt cluster coverage, indicating the formation of reduced iron oxide in vicinity of the clusters. Since magnetite is known to maintain its surface stoichiometry upon surface reduction by cation diffusion into the bulk,[55–57] those reduced FeO species must be located *on* the clusters, rather than around them, which is direct evidence for an SMSI effect. Note that due to the small cluster size, a stoichiometric, crystalline encapsulating layer, as observed for nanoparticles, is unlikely. The adsorption sites at the cluster surface are more likely blocked by a conglomerate of iron and oxygen atoms of unknown structure and stoichiometry that is hardly distinguishable from a Pt-FeO alloy.

To further confirm the encapsulation, we subsequently softly sputtered an encapsulated cluster sample, whereby indeed the encapsulating layer could be partially removed (restoring some of the original CO adsorption sites) and lost again during another TPD (see Supporting Information Figure S3). This particular encapsulation state is thus reproducibly obtained during a heating ramp to 820 K.

To sum up, the TPD experiments strongly indicate that the clusters get encapsulated by iron oxide upon annealing as a consequence of SMSI, whereby the desorption feature assignment indicates iron cations in a reduced state. We find that the loss in CO adsorption sites and hence cluster encapsulation also holds true for $Pt_5$ and $Pt_{10}$ (see Supporting Information Figure S4).

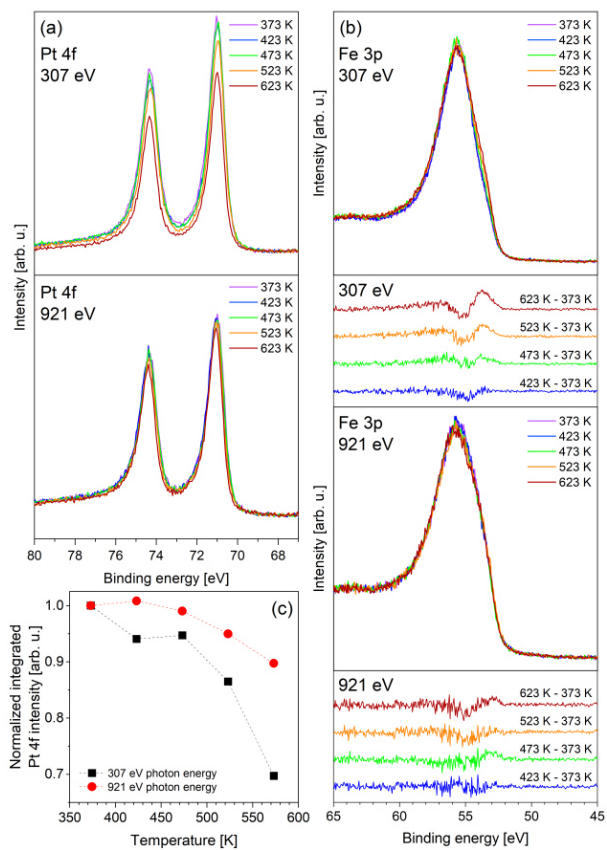

Figure 2. XP spectra of $Pt_{19}/Fe_3O_4(001)$ (0.05 clusters/nm$^2$) measured at 307 eV (highly surface-sensitive) and 921 eV photon energy (more bulk-sensitive), at the indicated temperatures in UHV. (a) Pt 4f spectra and (b) Fe 3p spectra with corresponding difference spectra for each photon energy below. (c) Evolution of the Pt 4f integrals for each photon energy, scaled to the first spectrum, respectively.

To confirm the presence of reduced FeO and the encapsulation of Pt clusters, we performed high resolution synchrotron XPS measurements on $Pt_{19}/Fe_3O_4(001)$, shown in Figure 2. We start by evaluating the attenuation of the Pt 4f signal induced by the encapsulating layer, comparing highly surface- and more bulk-sensitive measurements, recorded at 307 eV and 921 eV photon energy, respectively. Figure 2a displays the Pt 4f region measured at the indicated temperatures between 373 K and 623 K. The binding energy remains unchanged, but upon increasing the temperature, a significant decrease in intensity is observed starting around 523 K in the surface-sensitive spectra (top panel), whereas the more bulk-sensitive spectra change much less significantly (bottom panel). This effect is quantified in Figure 2c, where the XP spectra from Figure 2a are integrated and for each photon energy normalized to the corresponding 373 K spectrum. Evidently, the Pt 4f signal decrease is more pronounced for the surface-sensitive measurement, as expected for an SMSI-induced encapsulation of clusters, thus confirming our interpretation of the TPD experiments. A dissolution of the clusters into the bulk, on the other hand, can be excluded by the only slightly decreasing Pt 4f signal in the more bulk-sensitive measurement at 921 eV photon energy, as well as by the STM measurements (discussed below). This experiment already gives insights into the morphology of the encapsulating layer. Although alloying cannot be excluded entirely, the reduction of the signal intensity indicates an additional layer on top of the clusters, which is also in line with the sputtering experiment described above. Modeling the attenuation by an FeO encapsulating layer, a nominal average thickness of about 2.2 Å can be calculated,[37,58–61] comparable to the 2.5 Å for a bilayer in bulk FeO.[62] These findings agree well with those reported for larger Pt nanoparticles on $Fe_3O_4(111)$, where the encapsulating layer has been identified as an FeO bilayer under UHV conditions.[63]

Figure 2b shows the Fe 3p XPS region measured at the indicated temperatures. In the main peak, no significant change is observed for either photon energy, suggesting that the support overall remains unchanged upon heating. Cluster-related changes in the Fe 3p signal are expected to be minimal due to the low coverage. Indeed, we observe the development of a small low binding energy shoulder in the Fe 3p region measured with 307 eV photon energy, in the same temperature range in which the decrease of the Pt 4f signal occurs. Fitting these peaks is not trivial,[64,65] and minimal changes can be overlooked easily. We instead use the difference spectra for both photon energies to interpret the data, as shown in Figure 2b. Here, in the more surface-sensitive spectra the shoulder is apparent as a peak around 53.8 eV that increases in intensity when the temperature is increased beyond ~500 K. We assign this peak to $Fe^{2+}$.[64] At the same time, the signal corresponding to $Fe^{3+}$, at around 55.2 eV, decreases in intensity.[64] For the more bulk-sensitive measurement at 921 eV photon energy, the observed changes are in principle the same, but the intensity differences are significantly lower. Note that these transformations occur well below the known phase transition temperature of magnetite (001) around 720 K.[55] Thus, a slight increase in $Fe^{2+}$ content on the surface can be deduced, which we assign to the encapsulating SMSI layer, since all major changes in the Fe 3p signal develop around the same temperature where the highest decrease in the Pt 4f signal is observed. In agreement with our interpretation of the TPD experiments, the encapsulating layer on sub-nm clusters thus consists of reduced iron oxides, presumably FeO-like, similar to larger nanoparticles. The appearance of the encapsulating $Fe^{2+}$ species is not necessarily a direct transformation of surface-$Fe^{3+}$ species, but rather an extraction of interstitials to form the encapsulation layer. Indeed, DFT calculations showed that surface-$Fe^{3+}$ extraction and subsequent adsorption on a Pt cluster is rather energy intensive.[44]

Summarizing, our TPD and XPS experiments provide a first confirmation that encapsulation induced by SMSI indeed occurs also for sub-nanometer clusters and not only for extended nanoparticles and islands. We could identify the encapsulating layer as reduced defect-rich iron oxide, which covers the clusters.

*Cluster size-dependent sintering mechanisms.*

Having established encapsulation of Pt clusters, we now proceed to investigating how cluster ripening occurs under the SMSI conditions identified above. In particular, we investigate which ripening mechanisms occur, whether the cluster size plays a role, and what the sintered particles look like. $Pt_5$, $Pt_{10}$ and $Pt_{19}$ clusters on $Fe_3O_4(001)$ were deposited with similar cluster coverages at room temperature and subsequently annealed to increasingly higher temperatures in 50 K steps. Figure 3 shows representative STM images at the temperatures where the most significant changes occur (for a complete set of images see Supporting Information Figures S5 – S7), while a detailed height profile analysis of ten images at each temperature and cluster size is given in Figure 4 (the complete set of height histograms is shown in Figures S8 – S10). The clusters appear as bright protrusions in STM images, while the atomic-scale contrast and defects of the magnetite surface are also visible with a lower corrugation. We investigate four main temperature ranges in turn below, which we will relate to as-deposited size-selected clusters, lattice oxygen reverse spillover and encapsulation, cluster diffusion (Smoluchowski ripening), and Ostwald ripening.

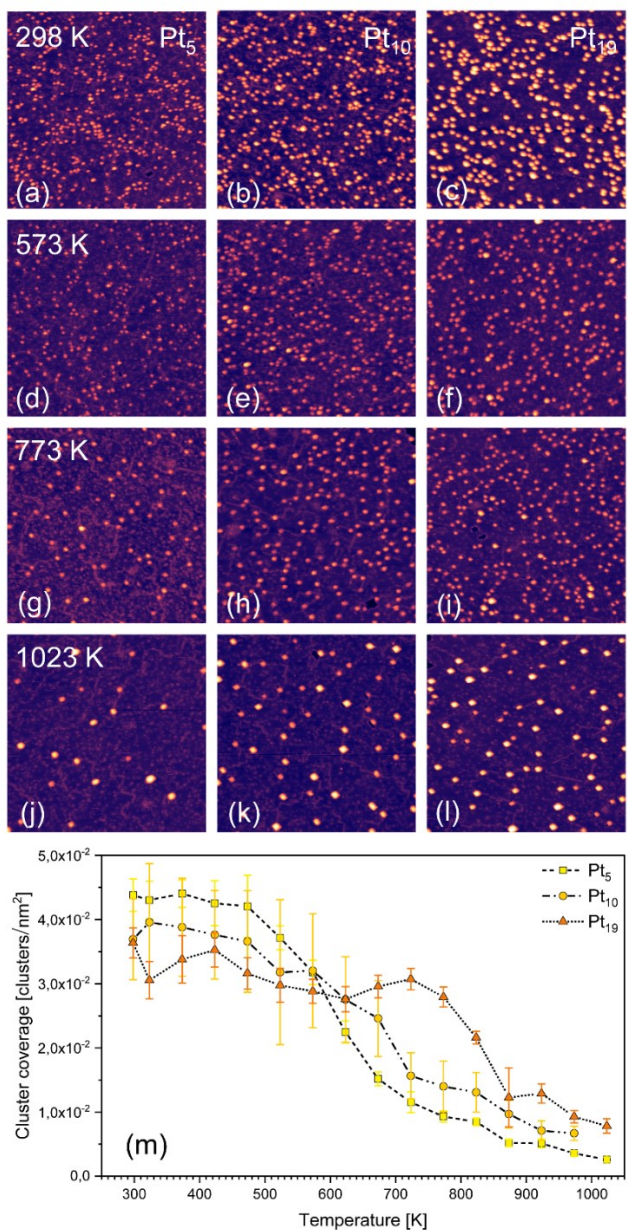

Figure 3: STM images of $Pt_5$ (left), $Pt_{10}$ (center) and $Pt_{19}$ (right column) clusters on $Fe_3O_4$(001) (~0.05 clusters/nm$^2$) are shown (a-c) as-deposited, after annealing for 10 min each to (d-f) 573 K, (g-i) 773 K and (j-l) 1023 K. For height comparison, all images use the same color scale. *Imaging conditions:* 1.5 V, 300 pA, 100 x 100 nm$^2$, RT. (m) Cluster coverage as a function of temperature, extracted from ten 100 x 100 nm$^2$ images for each condition and cluster size, leading to an initial number of approx. 5000 clusters contributing to the evaluation.

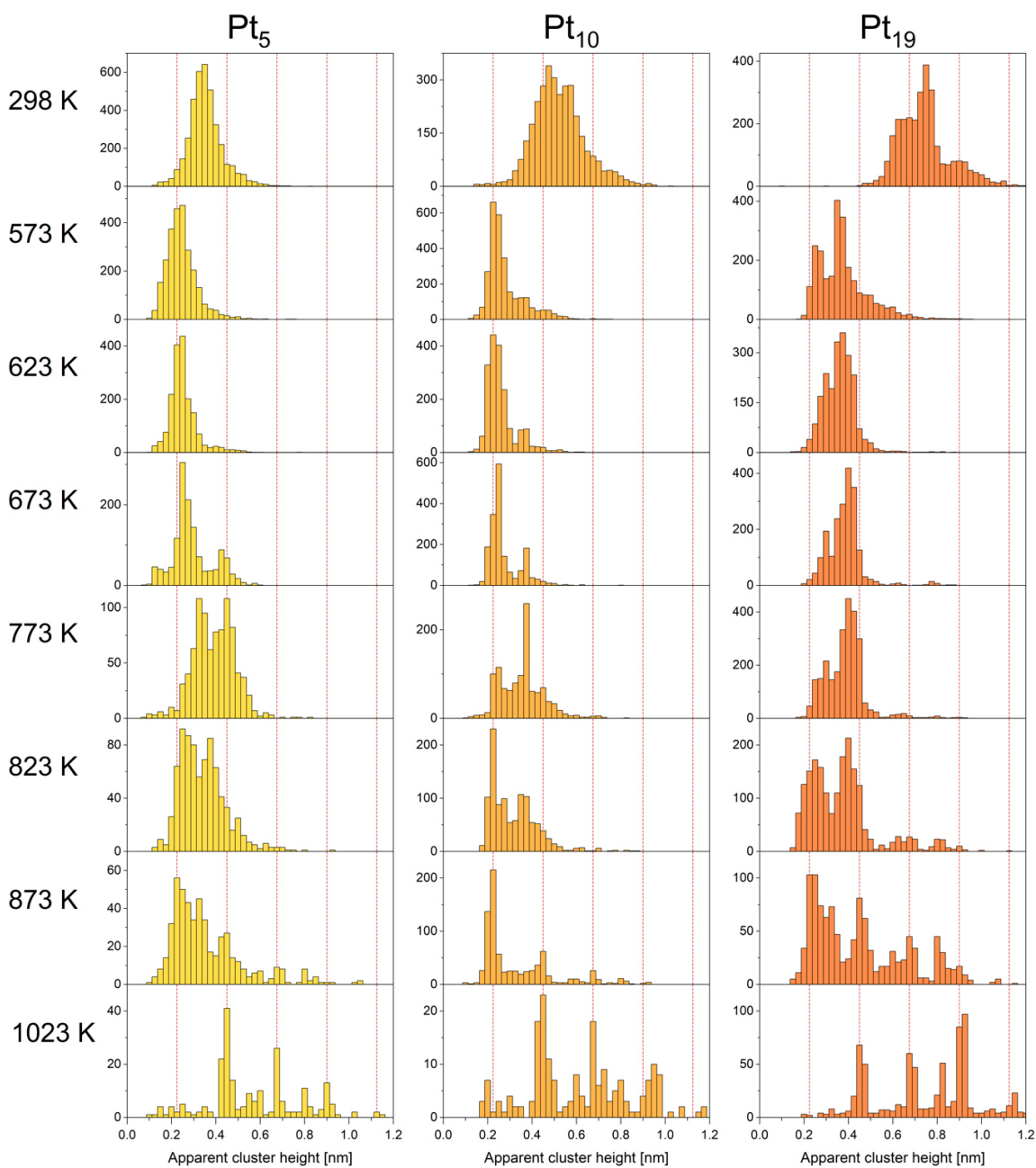

Figure 4: Histograms showing height distributions of $Pt_5$, $Pt_{10}$, and $Pt_{19}$ clusters on $Fe_3O_4(001)$ for different selected annealing temperatures. Absolute cluster numbers are indicated on the y-scale. For each temperature and cluster size, an area of 100 000 $nm^2$ was analyzed (corresponding to the same data used in Figure 3m). To guide the eye, we added red dotted lines in bulk Pt(111) step increments.

As evidenced by the STM images in Figure 3a-c and corresponding height profiles in Figure 4, the as-deposited clusters are indeed size-selected, intact and randomly distributed on the surface. To give a reference for cluster layering, we can use the height distributions and compare them with the bulk Pt(111) step height of about 2.26 Å,[66] while keeping in mind that clusters typically have non-fcc structures and their apparent heights are influenced by electronic effects. The as-deposited $Pt_5$ clusters exhibit a height distribution with a single clear peak between one and two atomic layers. In contrast, the distribution for $Pt_{10}$ clusters has a double peak around two and 2.5 layers and $Pt_{19}$ clearly displays at least three separate heights, with a most common apparent height between three and four atomic layers. The different apparent cluster heights within each cluster size distribution are most likely due to structural isomers and various adsorption sites on the surface.

When annealing up to 423 K, no significant changes are observed in the coverage nor height profiles for all three cluster sizes. In the temperature window from 473 K to 573 K (see Supporting Information), a decrease in apparent height is observed for all cluster sizes, which reaches a minimum after annealing to 623 K, yielding approx. single layer $Pt_5$ and $Pt_{10}$ clusters and $Pt_{19}$ between one and two layers. Note that the cluster coverage initially remains unchanged but begins to drop for $Pt_5$ and $Pt_{10}$ due to cluster diffusion-induced sintering at cluster footprint-dependent onset temperatures, as seen in Figures 3m and S8 – S10. As we described previously, the initial height decrease of the clusters results from lattice oxygen reverse spillover.[44] Furthermore, as shown above, the clusters become encapsulated by the reduced FeO layer in the same temperature range where a semiconductor-like conductivity has been described for encapsulated Pt nanoparticles on $TiO_2(110)$.[4] Such a change in the electronic nature of the particle can additionally affect the imaging and therefore also contribute to the observed decrease in apparent height.

For $Pt_5$ and $Pt_{10}$ clusters, the apparent cluster height increases again upon further annealing to 773 K, coinciding with a pronounced decrease in coverage due to sintering. With each temperature step, the coverage decreases further, until 773 K, where the curve flattens again. At the same time, the coverage and height distribution for $Pt_{19}$ clusters remains unchanged, i.e. the clusters are still size-selected. Notably, for all three initial cluster sizes, the height distributions now have a similar width predominantly between one and 2.5 layers. The sinter resistance of the larger $Pt_{19}$ combined with the similar resulting height distributions for all three initial sizes are clear evidence for Smoluchowski ripening, i.e. sintering by cluster diffusion. The diffusion barrier for clusters scales with their footprint and thus their interaction with the support.[24] Our data indicates that all $Pt_{19}$ isomers have a footprint large enough to be stabilized against diffusion in this temperature regime, which is why the sintering process of the smaller clusters naturally terminates when all clusters reach this size range. Therefore, already the coalescence of two $Pt_{10}$ or four $Pt_5$ clusters is sufficient to immobilize the resulting particles which makes the formation of larger nanoparticles statistically unlikely. Consequently, a narrow size distribution without much larger particles at this point is expected even after sintering, which is exactly what is observed in Figure 3g-i and Figure 4. As the coverage curve in Figure 3m demonstrates, the $Pt_5$ clusters begin to sinter at 523 K, i.e. 100 K before $Pt_{10}$, another strong indication for footprint dependence. Remarkably, we thus observe that even encapsulated clusters can still diffuse across the surface, confirming our hypothesis that an Fe-O agglomerate covers the clusters rather than a stoichiometric, rigid film.

By further increasing the temperature to 823 K, the situation changes. For all three investigated cluster samples, of which only the $Pt_{19}$ clusters were still size-selected up to 773 K, a new peak at lower heights of about one atomic layer appears in Figure 4, indicating a relatively larger proportion of smaller clusters. When annealing further to 873 K, this peak dominates the spectrum.

Additionally, weak peaks at three and four atomic layers start to appear, implying the formation of larger particles. This observation of diverging cluster sizes is typical for Ostwald ripening, where larger particles grow at the expense of smaller ones that exhibit a higher vapor pressure.[22] Further increasing the temperature to 1023 K, the now strongly sintered particles are mainly two to four atomic layers in height, with a small number being one or five layers high. At the same time, the coverage has decreased yet again for each initial cluster size. The particles formed at this final temperature adopt a square shape aligned with the cubic symmetry of the underlying support (Figure 3 and details enlarged in Figure S11) and have sharp apparent heights which can be clearly assigned to atomic layers, comparable to the results observed for atomic Pt deposition and sintering on the same support.[54] We enter a size regime where we start to have a clear crystalline-like layering while the internal atom arrangement and thus the overall particle shape are still dominated by the support registry. From the resulting particle coverage and size distribution, and taking into account the initially deposited amount of Pt, we roughly estimate that most particles contain between 50 and 150 atoms.

We have thus shown that both Smoluchowski and Ostwald ripening occur for size-selected Pt clusters on $Fe_3O_4$(001), whereby the mechanism strongly depends on the cluster size. Smaller clusters sinter via coalescence up to a size where the increasing diffusion barrier starts to be larger than those for atom detachment and Ostwald ripening dominates. Theoretical calculations for a comparable system, namely Pd clusters on $CeO_2$(111) indicate a similar effect.[24] There, Smoluchowski ripening is only possible for $Pd_4$ or smaller, whereas larger clusters sinter via Ostwald ripening.

As mentioned in the introduction, sintering of clusters via Ostwald ripening invariably begins at the empirical Hüttig temperature at the latest,[1,27] which is related to the melting point of the respective particle material and has been reported to be 608 K for Pt.[20] The actual onset temperature depends on the interaction with the support that influences the atom detachment energy. For unselected Pt clusters on $Si_3N_4$ films, for example, it has been shown that cluster sintering already sets in at 453 K.[25] In contrast, we observe Ostwald ripening of Pt on $Fe_3O_4$(001) starting at a temperature of 823 K, which is roughly 200 K *above* the Hüttig temperature, indicating a strong stabilization against sintering beyond the thermodynamic limit. With the strong experimental evidence for an encapsulating SMSI layer described above, we conclude that the clusters are stabilized by encapsulation. While the encapsulating layer clearly mitigates Ostwald ripening, the smallest clusters can still diffuse even when covered with an FeO-like conglomerate, leading to Smoluchowski ripening. The observation of moving encapsulated particles is in contrast to the recent literature, where an SMSI layer stabilizes Pt nanoparticles on titania against migration.[13]

4. CONCLUSION

In the present work, we have investigated the stability of small size-selected Pt clusters deposited on $Fe_3O_4$(001) in order to elucidate the strong metal-support interaction (SMSI) for catalytically active particles in the non-scalable size regime. Combined XPS and TPD experiments give clear evidence for the occurrence of encapsulation of sub-nanometer clusters. While the cluster surface is no longer accessible for CO adsorption, increased desorption from reduced iron oxide species points towards the formation of a defective FeO-like species encapsulating the clusters. Indeed, we observe the formation of a cluster-related $Fe^{2+}$ species, concomitant to an attenuation of the Pt 4f signal.

For a precise analysis of the cluster stability, we performed STM studies of $Pt_5$, $Pt_{10}$ and $Pt_{19}$, starting from comparable cluster coverages and annealing stepwise to 1023 K. The evolution of cluster coverage and apparent height indicates successive cluster coalescence (Smoluchowski) and Ostwald ripening regimes, with a transition temperature that strongly depends on the cluster footprint and thus on the initial cluster size. The final particle distribution appears to be independent from the initial cluster size and overall deposited Pt atom coverage, and shows square particle shapes with well-defined layer heights.

Remarkably, both sintering mechanisms are observed despite the encapsulation. That being said, Ostwald ripening sets in at an unexpectedly high temperature of 823 K, *i.e.* 200 K above the Hüttig temperature. We therefore conclude that the encapsulating layer strongly stabilizes the clusters against atom detachment, while still initially allowing diffusion of the smallest clusters. All our observations point to an encapsulating layer that is rather a conglomerate of Fe and O atoms surrounding the clusters than a stoichiometric, rigid iron oxide layer – as one might expect for particles close to the atomic limit of the size regime. It remains to be shown how the encapsulating layer changes in composition and morphology while clusters diffuse and coalesce, decay and grow.

ASSOCIATED CONTENT

**Supporting Information**. A pdf file containing the following information is available free of charge: S1. XP spectra including as-deposited measurements and desorption of adsorbates; S2. Additional $Pt_{19}$ deposition on already encapsulated $Pt_{19}/Fe_3O_4(001)$; S3. Partial removal of SMSI induced layer by soft sputtering; S4. Subsequent CO TPDs for $Pt_5$ and $Pt_{10}$ clusters on $Fe_3O_4(001)$; S5. Representative STM images for all annealing temperatures for $Pt_5$, $Pt_{10}$ and $Pt_{19}$ clusters on $Fe_3O_4(001)$; S6. Cluster height distributions for all annealing temperatures for $Pt_5$, $Pt_{10}$ and $Pt_{19}$ clusters on $Fe_3O_4(001)$; S7. Sintered Pt nanoparticles on $Fe_3O_4(001)$.

AUTHOR INFORMATION

**Corresponding Author**

* friedrich.esch@tum.de; bajlechner@tum.de

**Author Contributions**

‡ These authors contributed equally.

ACKNOWLEDGMENT

This work was funded by the Deutsche Forschungsgemeinschaft (DFG, German Research Foundation) under Germany's Excellence Strategy EXC 2089/1-390776260, through the project CRC1441 (project number 426888090), as well as by the grants ES 349/5-2 and HE 3454/ 23-2. It received funding from the European Research Council (ERC) under the European Union's Horizon 2020 research and innovation program (grant agreement no. 850764). B.A.J.L. gratefully acknowledges financial support from the Young Academy of Sciences and Humanities. We acknowledge MAX IV Laboratory for time on HIPPIE Beamline under Proposal 20200272. Research conducted at MAX IV, a Swedish national user facility, is supported by the Swedish Research council under contract 2018-07152, the Swedish Governmental Agency for Innovation Systems under contract 2018-04969, and Formas under contract 2019-02496.

# Supporting Information for

# Does cluster encapsulation inhibit sintering? Stabilization of size-selected Pt clusters on Fe$_3$O$_4$(001) by SMSI

*Sebastian Kaiser,[‡,1,2] Johanna Plansky,[‡,2,3] Matthias Krinninger,[2,3] Andrey Shavorskiy,[4] Suyun Zhu,[4] Ueli Heiz,[1,2] Friedrich Esch,[1,2,*] and Barbara A. J. Lechner [2,3,5,*]*

[1] Chair of Physical Chemistry, Department of Chemistry, School of Natural Sciences, Technical University of Munich, Lichtenbergstr. 4, 85748 Garching, Germany

[2] Catalysis Research Center, Technical University of Munich, Lichtenbergstr. 4, 85748 Garching, Germany

[3] Functional Nanomaterials Group, Department of Chemistry, School of Natural Sciences, Technical University of Munich, Lichtenbergstr. 4, 85748 Garching, Germany

[4] MAX IV Laboratory, Lund University, Lund 221 00 Sweden

[5] Institute for Advanced Study, Technical University of Munich, Lichtenbergstr. 2a, 85748 Garching, Germany

## S1. XP spectra including as-deposited measurements and desorption of adsorbates

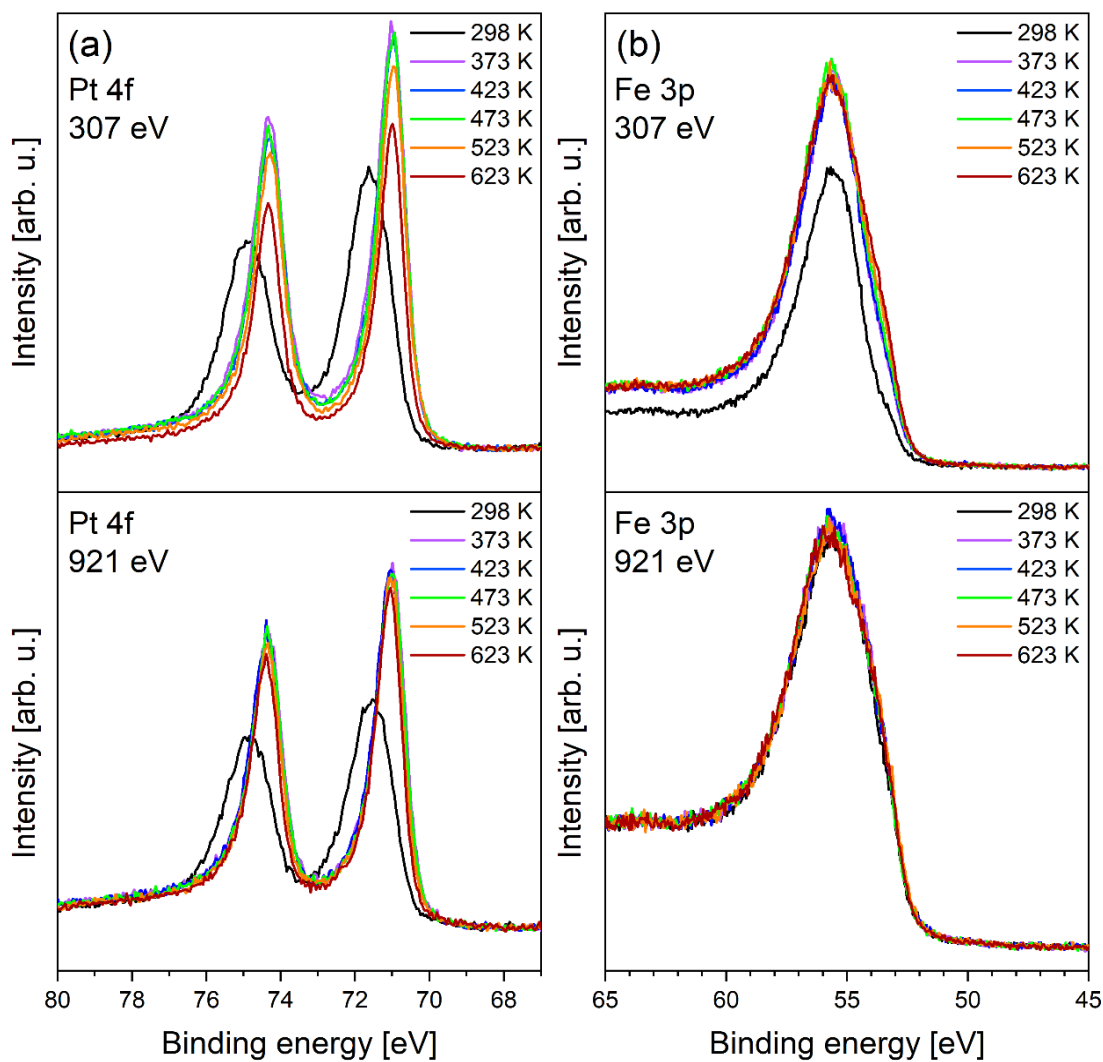

Figure S1. XP spectra including the as-deposited room temperature measurements of $Pt_{19}/Fe_3O_4(001)$ (0.05 clusters/nm$^2$) measured at 307 eV (upper panels, more surface sensitive) and 921 eV photon energy (lower panels, more bulk sensitive) at the indicated temperatures, showing the Pt 4f region (a), and Fe 3p region (b). To desorb any adsorbates accumulated during the transport in UHV (e.g. CO, H$_2$O or hydrocarbons), the sample has been initially heated to 373 K. A shift of 0.6 eV to lower binding energy as well as a reduction of the peak width is observed in the Pt 4f signal for both photon energies. This shift corresponds to the desorption of CO from the Pt clusters.[1,2] Simultaneously, the Fe 3p signal shows no clear shift, but a strong increase in signal intensity for 307 eV photon energy upon this first annealing step, which is not observed for the 921 eV measurement. This points to an adsorbate overlayer on top of the magnetite crystal, which desorbs upon annealing to 373 K. From there on, no large intensity changes are observed in the Fe 3p signal, implying that this first annealing is sufficient to remove the adsorbates.

## S2. Additional Pt₁₉ deposition on already encapsulated Pt₁₉/Fe₃O₄(001)

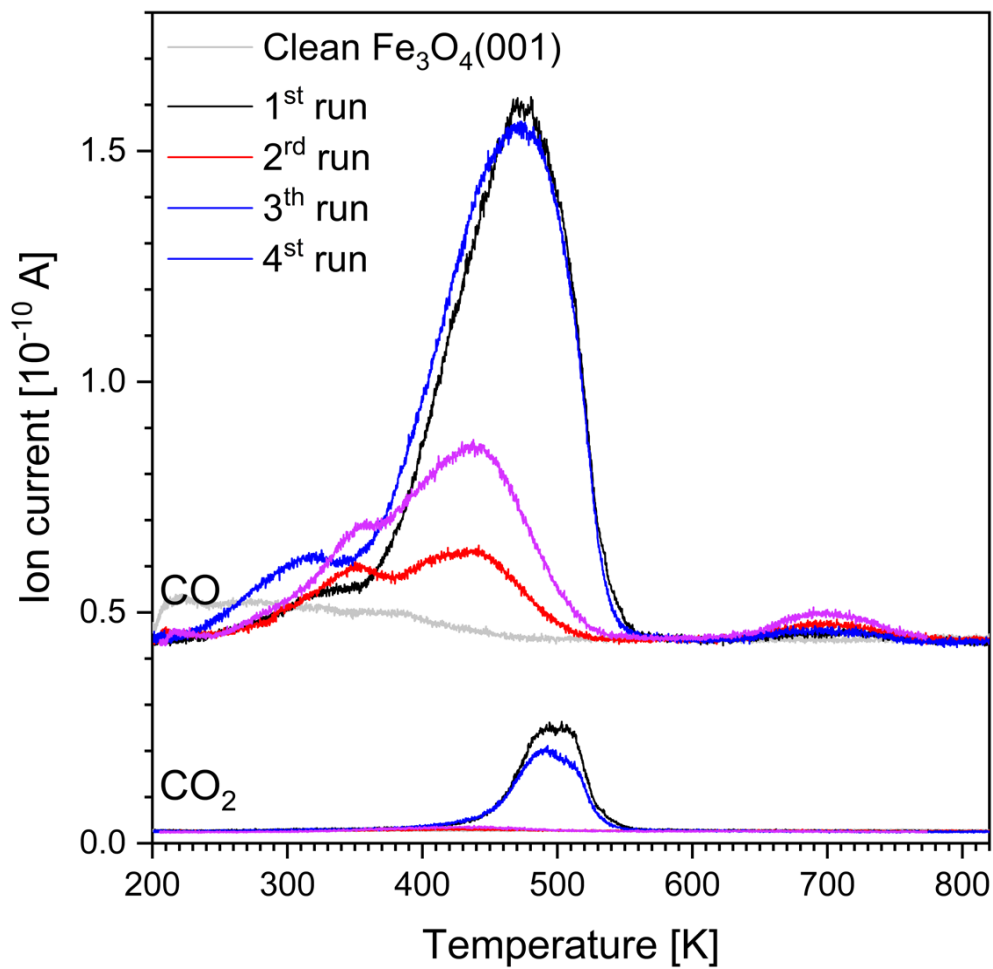

Figure S2. Subsequent TPD spectra (heating rate 1 K/s) of $Pt_{19}$ clusters on $Fe_3O_4(001)$ (initial coverage 0.05 clusters/nm$^2$), measured after saturating the sample with C$^{18}$O (~10 L) at 200 K. The C$^{18}$O (m/z = 30, top) as well as the C$^{18}$O$^{16}$O (m/z = 46, bottom) traces are shown. The desorption spectra from the bare $Fe_3O_4(001)$ substrate prior to cluster deposition are shown in gray. In the first run (black), a CO desorption feature corresponding to desorption from the clusters is observed at 475 K. CO$_2$ formation as a consequence of lattice oxygen reverse spillover is detected simultaneously.[3] In the second run (red), the cluster-related CO desorption as well as the CO$_2$ production have vanished completely. Two new features at 350 K and 430 K arise instead, corresponding to desorption from the encapsulating FeO-like conglomerate.[4] Upon deposition of another 0.05 clusters/nm$^2$ on the sample (blue), the initial peaks reappear in the CO and CO$_2$ traces, in addition to the second run features observed before. In a final, fourth run (magenta). the CO$_2$ formation has vanished once more, and again, only the encapsulating layer-related features are observed in the CO trace, with about twice the intensity of the second run. It becomes evident that the CO desorption from the FeO-like features directly scales with the cluster coverage and must therefore originate in the vicinity of the clusters. As the magnetite surface itself maintains its stoichiometry upon reduction by diffusion of undercoordinated iron into the bulk,[5,6] the desorption features appearing in the second and fourth runs confirm the formation of an encapsulating FeO-like layer on the clusters.

## S3. Partial removal of SMSI induced layer by soft sputtering

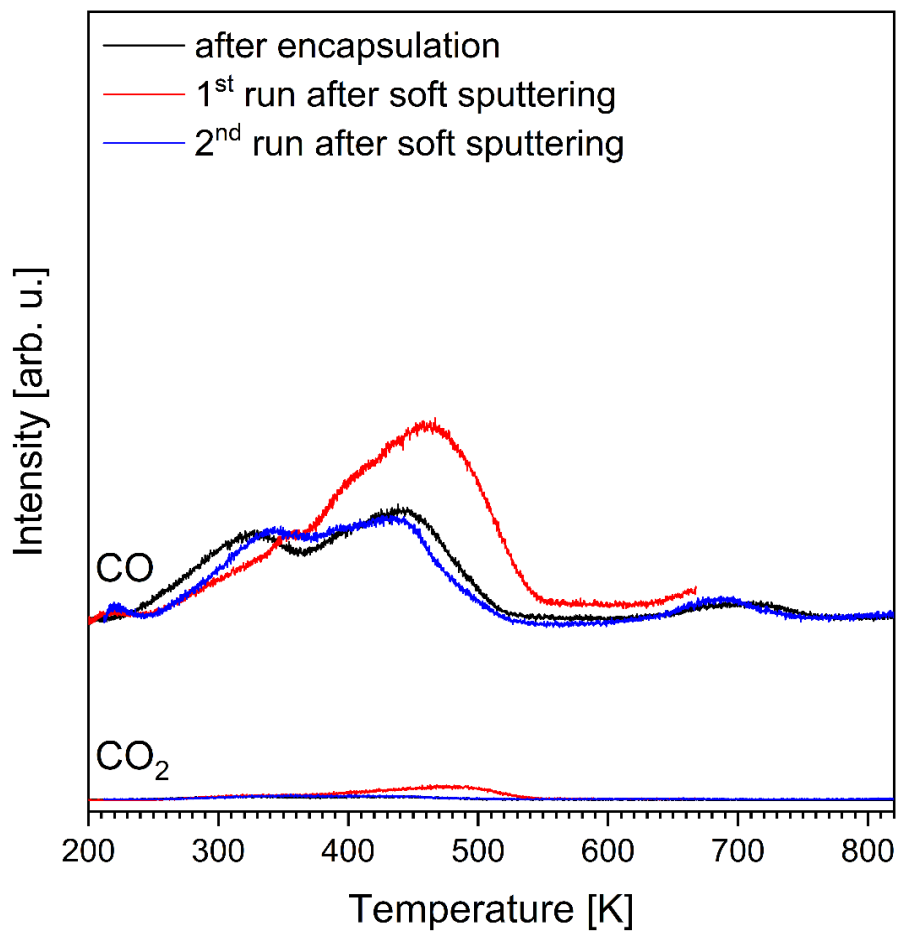

Figure S3. Subsequent TPD spectra (heating rate 1 K/s) of $Pt_{19}$ clusters on $Fe_3O_4(001)$ (initial coverage 0.05 clusters/nm$^2$), measured after saturating the sample with $C^{18}O$ (~10 L) at 200 K. Again, the $C^{18}O$ (m/z = 30, top) as well as the $C^{18}O^{16}O$ (m/z = 46, bottom) traces are shown. The clusters have been encapsualted as a consequence of SMSI by annealing to 820 K. The first run after encapsulation (black) shows the two characteristic CO desorption features from the FeO-like encapsulating conglomerate at 350 K and 430 K, while $CO_2$ formation is not observed. After softly sputtering the sample (40 s, 1 x 10$^{-6}$ mbar Ar, 1 keV), the CO desorption has changed (red). Here, the intensity of the SMSI layer-related 350 K feature decreases and a new peak arises with a maximum at 475 K. $CO_2$ formation is observed concurrently. The new peaks correspond to those observed for as-deposited clusters, giving clear evidence that the encapsulating layer could be partially removed by sputtering, recovering the pristine Pt cluster surface. It can thus be concluded that, upon annealing, the clusters become indeed encapsulated rather than forming an alloy. Note, that although the measurement has been stopped earlier, the sample was still annealed to 820 K in this run. A last TPD experiment (blue) is highly similar to the first run after encapsulation (black), showing esentially the same desorption features and intensities, meaning that after this soft sputtering, the clusters are re-encapsulated reproducibly. We thus conclude that during sputtering the majority of the clusters have stayed intact and just the encapsulating layer has been removed.

## S4. Subsequent CO TPDs for Pt$_5$ and Pt$_{10}$ clusters on Fe$_3$O$_4$(001)

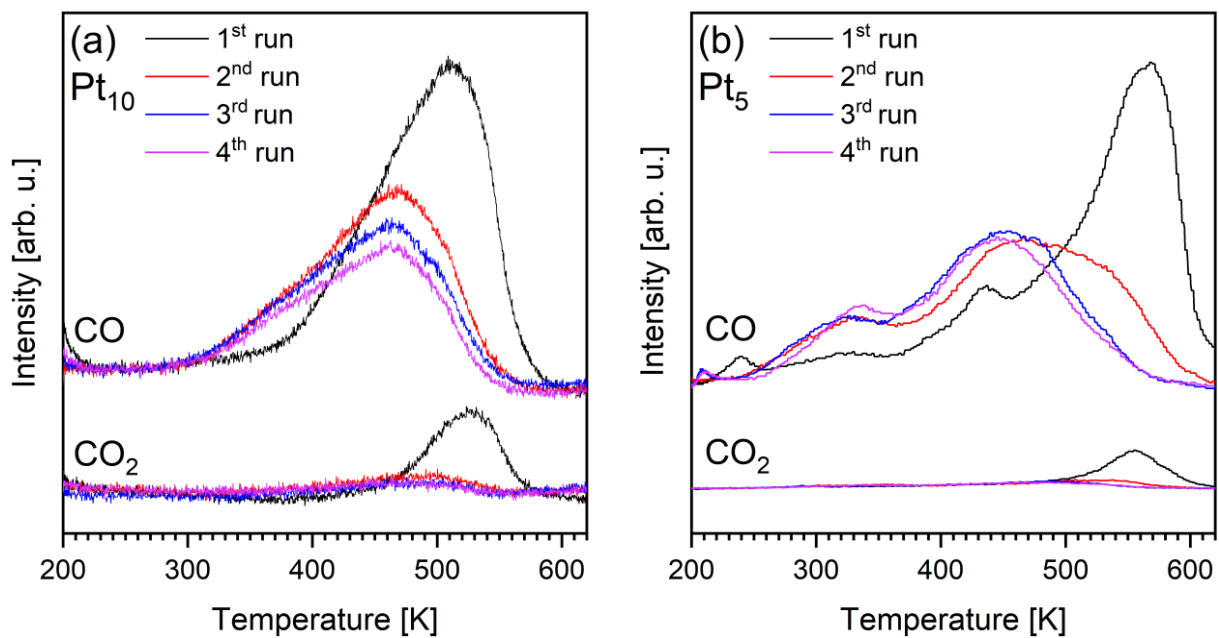

Figure S4. C$^{18}$O TPD spectra of (a) Pt$_{10}$ and (b) Pt$_5$ clusters deposited on an Fe$_3$O$_4$(001) surface (0.05 clusters/nm$^2$), showing the C$^{18}$O (m/z = 30, top) and the C$^{18}$O$^{16}$O (m/z = 46, bottom) traces. The samples have been saturated with CO at 200 K (~10 L) before each run, the heating rate was 1 K/s. Both cluster sizes show a clear desorption peak in the first run (black), at 510 K for Pt$_{10}$ and 550 K for Pt$_5$, which is attributed to the clusters. Analogous to Pt$_{19}$ (see Figure 1), CO$_2$ formation is observed concurrently with the main CO desorption features for both cluster sizes. In the second run (red), the main CO desorption feature is strongly reduced in intensity, while new peaks arise at 360 K and 460 K for Pt$_{10}$ and 325 K and 450 K, which are assigned to desorption from an encapsulating FeO-like layer. As expected, also the CO$_2$ formation has decreased significantly. When completing the encapsulating layer growth in the third (blue) and fourth (magenta) runs, the CO$_2$ formation, as well as the original cluster-related CO desorption peaks vanish completely. These results are very similar to those observed for Pt$_{19}$ clusters. We thus conclude that Pt$_5$ and Pt$_{10}$ clusters are also encapsulated by a reduced iron oxide conglomerate upon annealing.

**S5. Representative STM images for all annealing temperatures for Pt$_5$, Pt$_{10}$ and Pt$_{19}$ clusters on Fe$_3$O$_4$(001)**

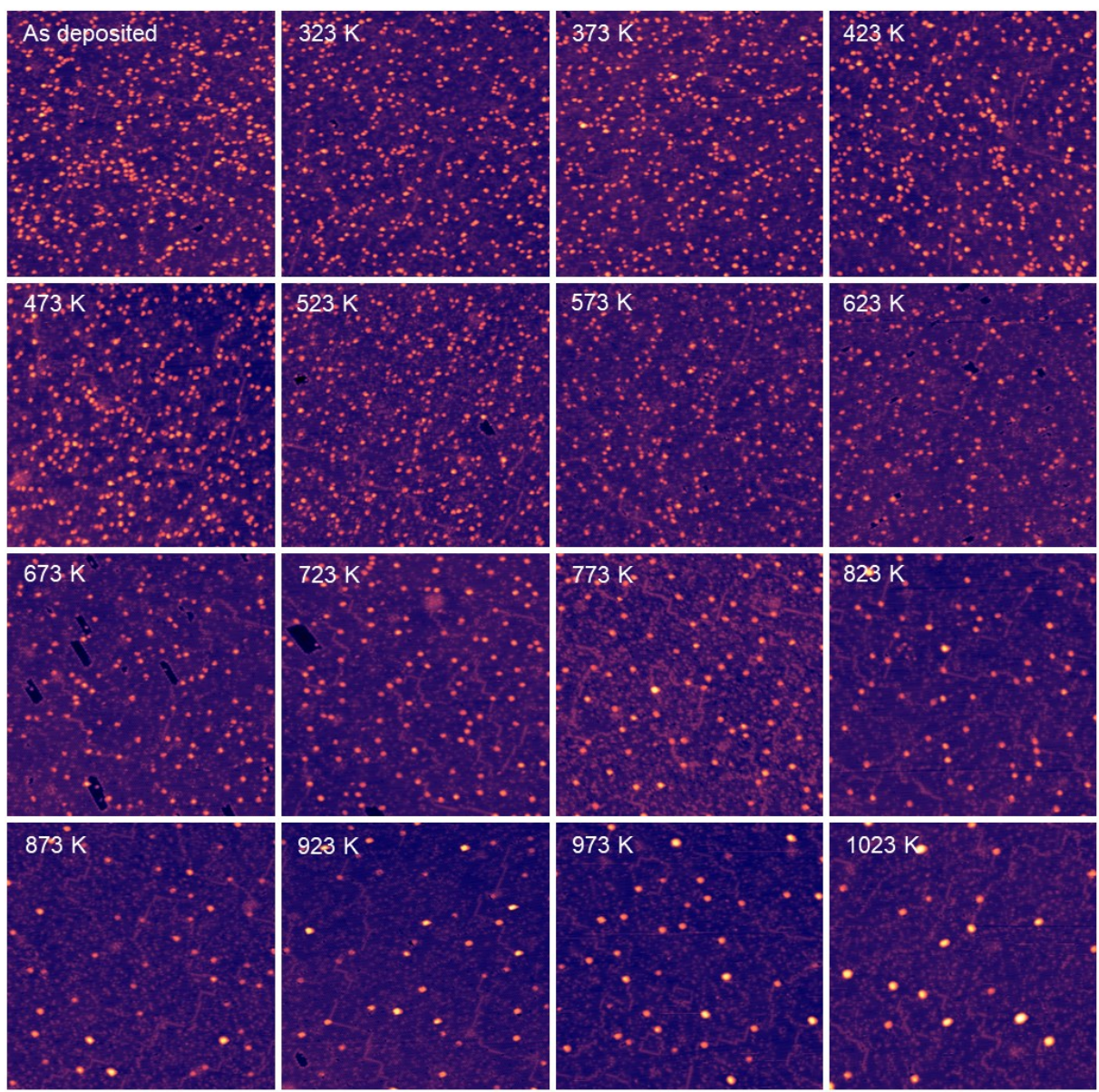

Figure S5. Representative set of STM images of Pt$_5$ clusters on Fe$_3$O$_4$(001) (initial coverage of 0.05 clusters/nm$^2$), measured at room temperature after annealing for 10 min to the temperatures indicated in the respective panels. For easier cluster height comparison, all Figures of Section S5 are set to the same color scale of 1.4 nm range. *Imaging conditions:* 1.5 V, 300 pA, 100 x 100 nm$^2$.

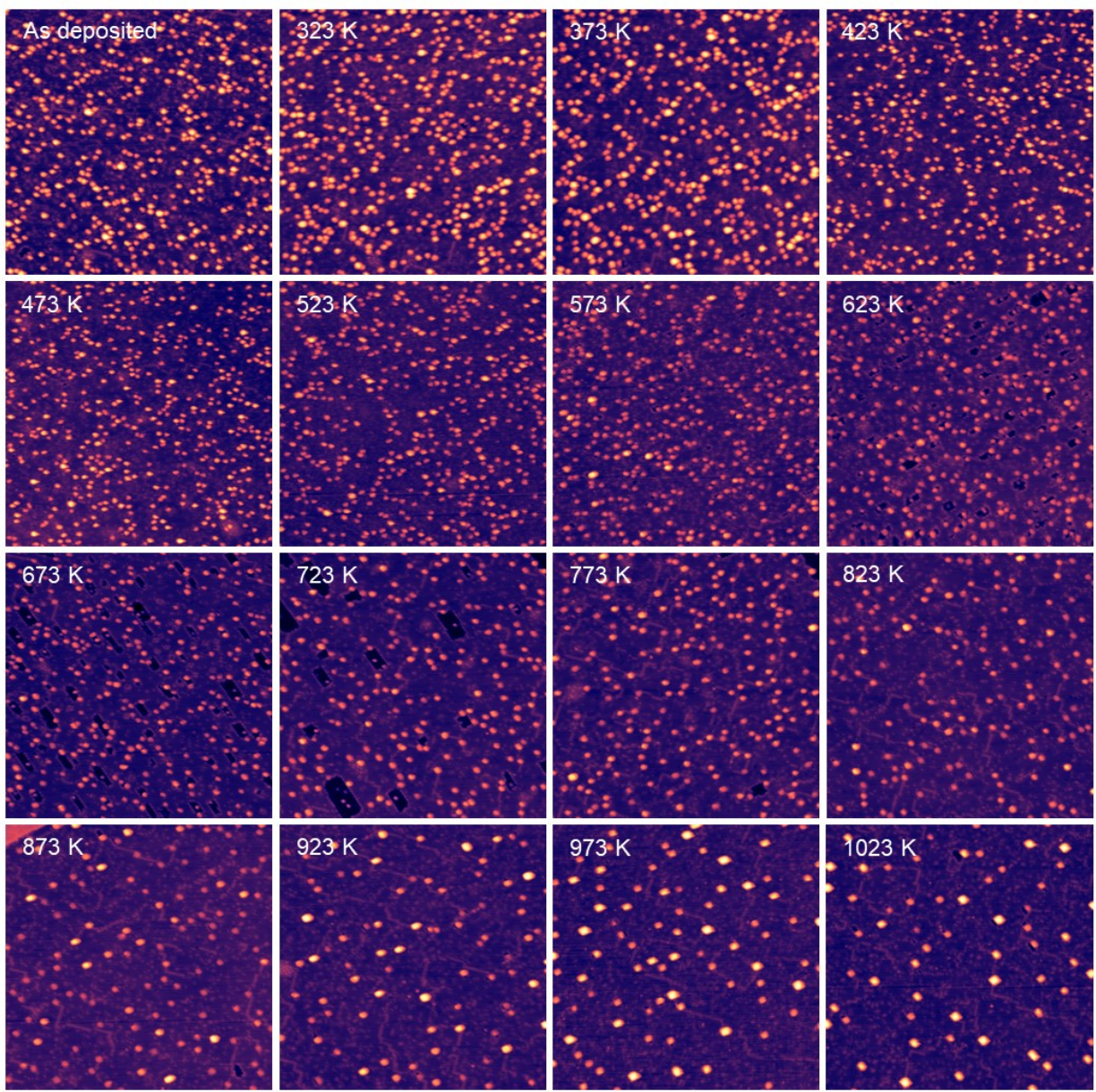

Figure S6. Representative set of STM images of $Pt_{10}$ clusters on $Fe_3O_4(001)$ (initial coverage of 0.05 clusters/nm$^2$), measured at room temperature after annealing for 10 min to the temperatures indicated in the respective panels. For easier cluster height comparison, all Figures of Section S5 are set to the same color scale of 1.4 nm range. *Imaging conditions:* 1.5 V, 300 pA, 100 x 100 nm$^2$.

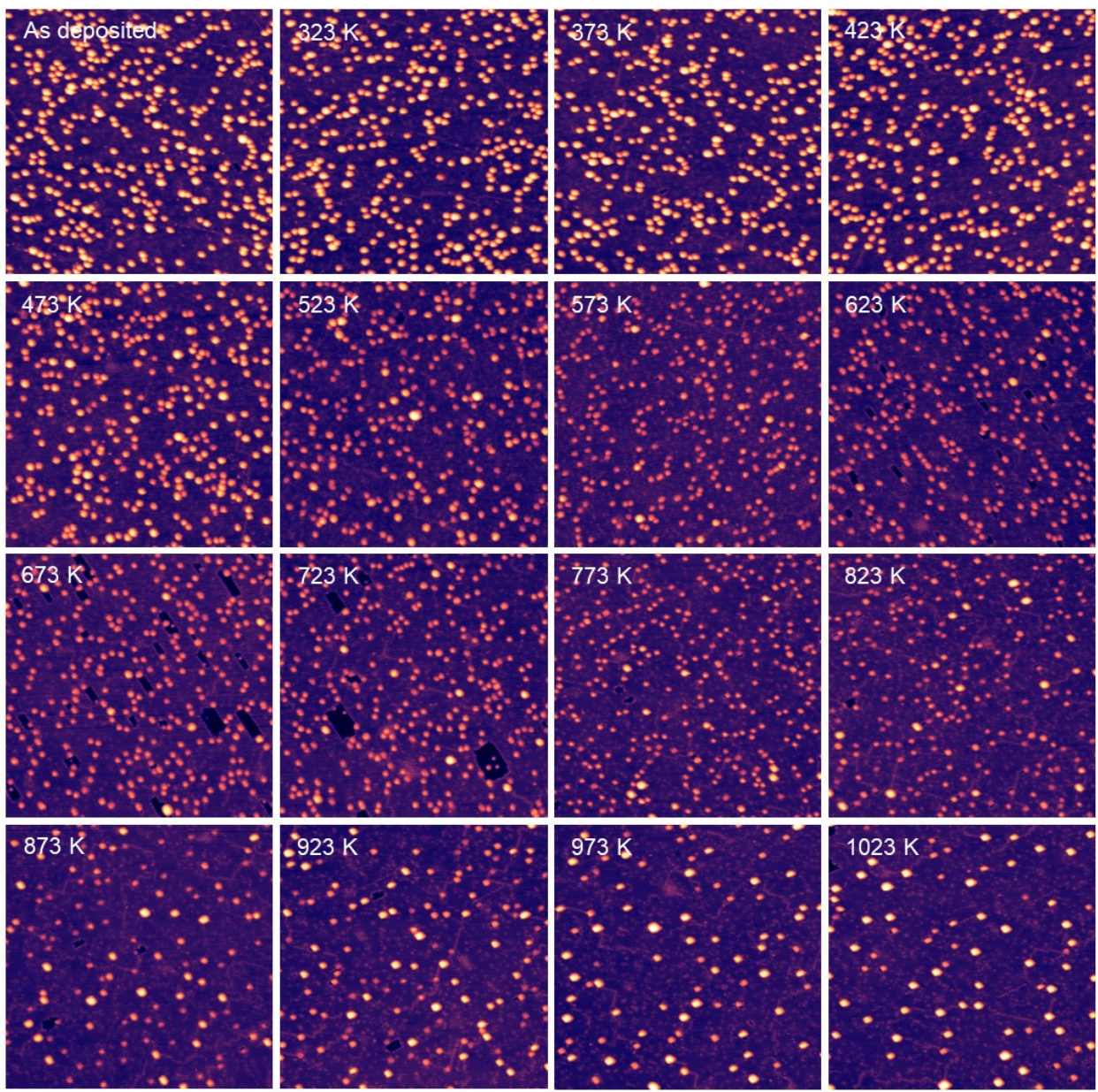

Figure S7. Representative set of STM images of $Pt_{19}$ clusters on $Fe_3O_4(001)$ (initial coverage of 0.05 clusters/nm$^2$), measured at room temperature after annealing for 10 min to the temperatures indicated in the respective panels. For easier cluster height comparison all Figures of Section S5 are set to the same color scale of 1.4 nm range. *Imaging conditions:* 1.5 V, 300 pA, 100 x 100 nm$^2$.

## S6. Cluster height distributions for all annealing temperatures for $Pt_5$, $Pt_{10}$ and $Pt_{19}$ clusters on $Fe_3O_4(001)$

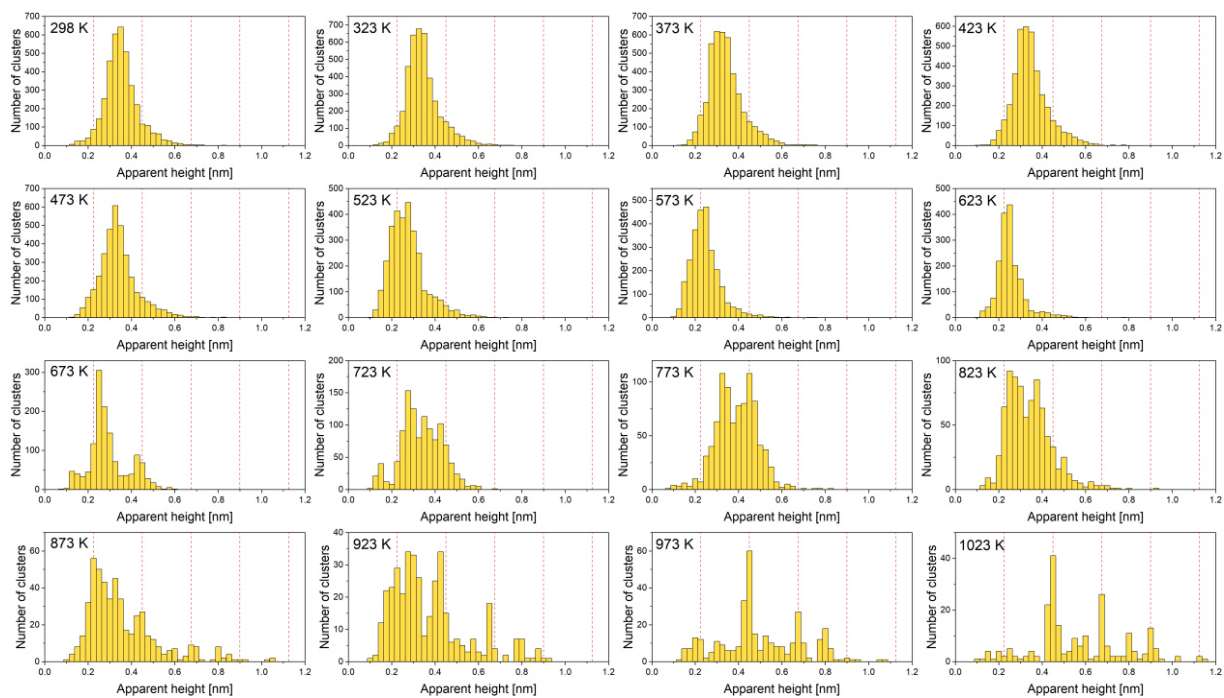

Figure S8. Apparent cluster height distributions of $Pt_5/Fe_3O_4(001)$ (initial coverage of 0.05 clusters/nm$^2$), corresponding to the data in Figure S5 and nine further images at each condition. A total area of 100 000 nm$^2$ has been investigated. The annealing temperatures are indicated in the respective graphs. Red dotted lines indicating the atomic step height of a Pt(111) surface are added as a guide to the eye.

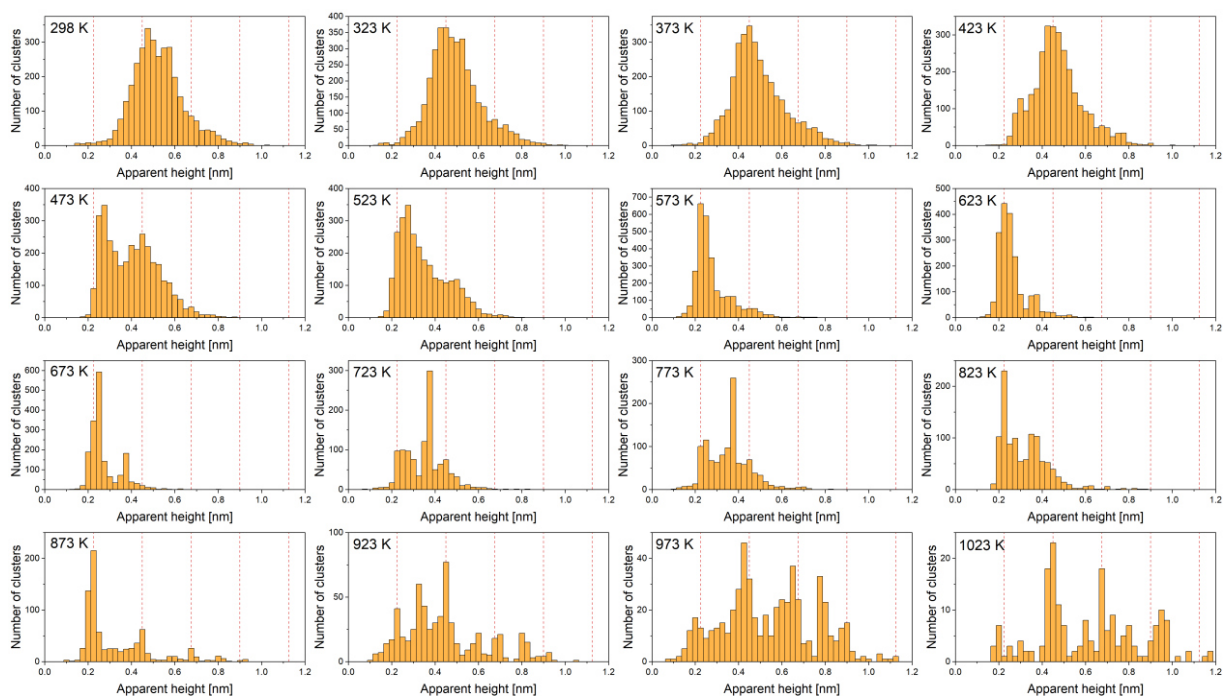

Figure S9. Apparent cluster height distributions of $Pt_{10}/Fe_3O_4(001)$ (initial coverage of 0.05 clusters/nm$^2$), corresponding to the data in Figure S6. and nine further images at each condition. A total area of 100 000 nm$^2$ has been investigated. The annealing temperatures are indicated in the respective graphs. Red dotted lines indicating the atomic step height of a Pt(111) surface are added as a guide to the eye. Note that the transition to single layer clusters between 423 K and 473 K sets in slightly earlier than for the $Pt_5$ sample. This is related to a lower CO desorption temperature on $Pt_{10}$, which is prerequisite for the lattice oxygen reverse spillover.[3]

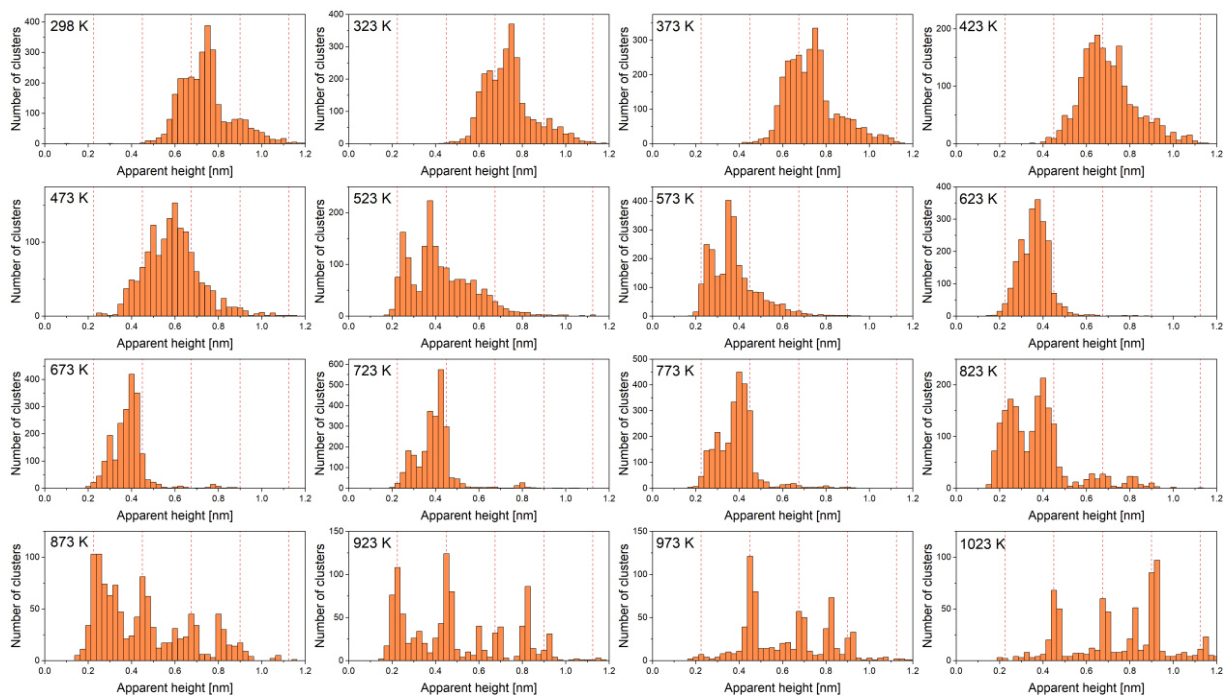

Figure S10. Apparent cluster height distributions of Pt$_{19}$/Fe$_3$O$_4$(001) (initial coverage of 0.05 clusters/nm$^2$), corresponding to the data in Figure S7. and nine further images at each condition. A total area of 100 000 nm$^2$ has been investigated. The annealing temperatures are indicated in the respective graphs. Red dotted lines indicating the atomic step height of a Pt(111) surface are added as a guide to the eye.

## S7. Sintered Pt nanoparticles on Fe$_3$O$_4$(001)

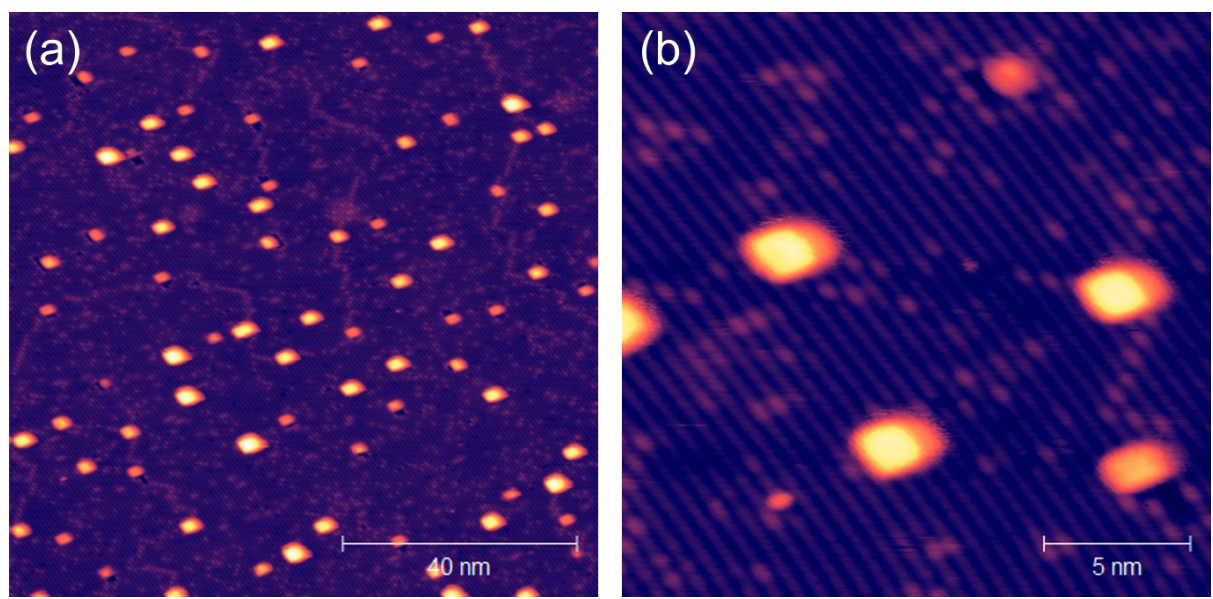

Figure S11. Pt nanoparticles on Fe$_3$O$_4$(001), grown by sintering of Pt$_{19}$ clusters at 1023 K for 10 min. (a) The overview image shows that all particles exhibit a rectangular, rather crystalline shape, adapting the cubic symmetry of the underlying magnetite substrate.[7] The particles are estimated to consist of roughly 50 – 150 atoms. (b) Zooming in reveals that all nanoparticles are aligned with the atomic rows of the substrate, confirming the particle growth being dominated by the cubic symmetry of the surface. *Imaging conditions:* 1.5 V, 300 pA, (a) 100 x 100 nm$^2$, (b) 20 x 20 nm$^2$.